\documentclass[lnbip]{svmultln}
\usepackage{graphicx} 
\usepackage{makeidx}  
\usepackage{amsmath}  
\usepackage{algorithm}
\usepackage{algcompatible}%

\algnewcommand\algorithmicreturn{\textbf{return}}
\algnewcommand\RETURN{ \State \algorithmicreturn }%

\begin{document}
\mainmatter              
\title{Urban Mobility Swarms: A Scalable Implementation}

\titlerunning{Urban Mobility Swarms}  
%
\author{Alex Berke \and Jason Nawyn \and Thomas Sanchez Lengeling \and Kent Larson}

\authorrunning{Alex Berke et al.}   

\tocauthor{Alex Berke, Jason Nawyn, Thomas Sanchez Lengeling, Kent Larson}

\institute{MIT Media Lab, Massachusetts Institute of Technology, Cambridge MA 02139, USA,\\
\email{{aberke}, {nawyn}, {thomassl}, {kll}@media.mit.edu},\\ WWW home page:
\texttt{https://www.media.mit.edu/groups/city-science/overview/}
}

\maketitle              
\index{Berke, Alex}   
\index{Nawyn, Jason}  
\index{Lengeling, Thomas Sanchez}
\index{Larson, Kent}

\begin{abstract}        
We present a system to coordinate ``urban mobility swarms'' in order to promote the use and safety of lightweight, sustainable transit, while enhancing the vibrancy and community fabric of cities.  This work draws from behavior exhibited by swarms of nocturnal insects, such as crickets and fireflies, whereby synchrony unifies individuals in a decentralized network. Coordination naturally emerges in these cases and provides a compelling demonstration of ``strength in numbers''.  Our work is applied to coordinating lightweight vehicles, such as bicycles, which are automatically inducted into ad-hoc ``swarms'', united by the synchronous pulsation of light. We model individual riders as nodes in a decentralized network and synchronize their behavior via a peer-to-peer message protocol and algorithm, which preserves individual privacy. Nodes broadcast over radio with a transmission range tuned to localize swarm membership.  Nodes then join or disconnect from others based on proximity, accommodating the dynamically changing topology of urban mobility networks. This paper provides a technical description of our system, including the protocol and algorithm to coordinate the swarming behavior that emerges from it.  We also demonstrate its implementation in code, circuity, and hardware, with a system prototype tested on a city bike-share.  In doing so, we evince the scalability of our system.  Our prototype uses low-cost components, and bike-share programs, which manage bicycle fleets distributed across cities, could deploy the system at city-scale.  Our flexible, decentralized design allows additional bikes to then connect with the network, enhancing its scale and impact.
\end{abstract}

\section{Introduction}
Cities comprise a variety of mobility networks, from streets and bicycle lanes, to rail and highways. Increasing the use of the lightweight transit options that navigate these networks, such as bicycles and scooters, can increase the sustainability of cities and public health \cite{de2010health} \cite{johansson2017impacts} \cite{bbc_news_2016_air_pollution}.  However, infrastructure to promote and protect lightweight transit, such as bicycle lanes, are limited, and riders are vulnerable on streets designed to prioritize the efficient movement of heavier vehicles, such as cars and trucks.

In this paper we present our design and implementation of a system that synchronizes lights of nearby bicycles, automatically inducting riders into unified groups (swarms), to increase their presence and collective safety.  Ad-hoc swarms emerge from our system, in a distributed  network that is superimposed on the physical infrastructure of existing mobility networks.  We designed and tested our system with bicycles, but our work can be extended to unify swarms of the other lightweight and sustainable transit alternatives present in cities.

As bicycles navigate dark city streets, they are often equipped with lights.  The lights are to make their presence known to cars or other bikers, and make the hazards of traffic less dangerous.  As solitary bikes equipped with our system come together, their lights begin to softly pulsate, at the same cadence.  The cyclists may not know each other, or may only pass each other briefly, but for the moments they are together, their lights synchronize.  The effect is a visually united presence, as swarms of bikes illuminate themselves with a gently breathing, collective light source.  As swarms grow, their visual effect and ability to attract more cyclists is enhanced.  The swarming behavior that results is coordinated by our system technology without effort from cyclists, as they collaboratively improve their aggregate presence and safety.

We provide a technical description of our light system that includes the design of a peer-to-peer message protocol, algorithm, and low-cost hardware.  We also present our prototypes that were tested on a city bike-share network.  The system's low-cost and the opportunity for bike-share programs to deploy it city-wide allows the network of swarms to quickly scale. In addition, the decentralized and flexible nature of our design allows new bikes to join a network, immediately coordinate with other bikes, and further grow a network of swarms.

Our system is designed for deployment in a city, yet draws inspiration from nature.  Swarms of insects provide rich examples of synchrony unifying groups of individuals in a decentralized network.  We focus on examples particular to the night. 

The sound of crickets in the night is the sound of many individual insects, chirping in synchrony.  A single cricket's sound is amplified when it joins the collective whole. The spectacle of thousands of male fireflies gathering in trees in southeast Asia to flash in unison has long been recorded and studied by biologists \cite{buck1938synchronous}, \cite{buck1976synchronous}.

These examples of synchrony emerging via peer-to-peer coordination within a decentralized network are of interest in our design for urban swarms.  They have also interested biologists, who have studied the coordination mechanisms of these organisms \cite{strogatz1993coupled}.  Applied mathematicians and physicists have also analyzed these systems and attempted to model the dynamics of their synchronized behavior \cite{mirollo1990synchronization}, \cite{werner2005firefly}.  We draw from these prior technical descriptions in order to describe the coordination of our decentralized bike light system.

In doing so, we describe the individual bikes that create and join swarms as nodes in a distributed network.  These nodes are programmed to behave as oscillators, and their synchronization is coordinated by aligning their phases of oscillation via exchange of peer-to-peer messages.  Our message protocol and underlying algorithm accommodate the dynamically changing nature of urban mobility networks. New nodes can join the network, and nodes can drop out, and yet our system maintains its mechanisms of synchrony. Moreover, our system is scalable due to its simplicity, flexibility, and features that allow nodes to enter the swarm network with minimal information and hardware.  Namely:

\begin{itemize}
\item There is no global clock
\item Nodes communicate peer-to-peer via simple radio messages
\item Nodes need not be predetermined nor share metadata about their identities
\item Nodes can immediately synchronize
\end{itemize}

Before we provide our technical description and implementation, we first describe the bicycles, their lights, and their swarming behavior. We then describe them as nodes in a dynamic, decentralized network of swarms, before presenting our protocol and algorithm that coordinates their behavior. Lastly, we show how we prototyped and tested our system with bicycles from a city bike-share program.

\section{Swarm Behavior and Bicycles Lights}
Similar to our swarms of bicycles, swarms of nocturnal insects, such as crickets and fireflies, display synchronous behavior within decentralized networks. In these cases, the recruitment and coordination of individuals in close proximity emerges from natural processes and provides a compelling demonstration of ``strength in numbers.''

This concept of ``strength in numbers'' demonstrated in natural environments can be extended to the concept of ``safety in numbers'' for urban environments.  ``Safety in numbers'' is the hypothesis that individuals within groups are less likely to fall victim to traffic mishaps, and its effect has been well studied and documented in bicycle safety literature \cite{/content/publication/9789282105955-en} \cite{jacobsen2015safety}. The cyclists within swarms coordinated by our system are safer due to their surrounding numbers, but also because their presence is pronounced by the visual effect swarms produce with their synchronized lights.

Unlike insects, the coordination of bikes swarms is due to peer-to-peer radio messages and software, yet swarms can still form organically when cyclists are in proximity.  The visual display of synchronization is due to the oscillating amplitude of LED lights.

\begin{figure}
\centering
\includegraphics[width=0.5\columnwidth]{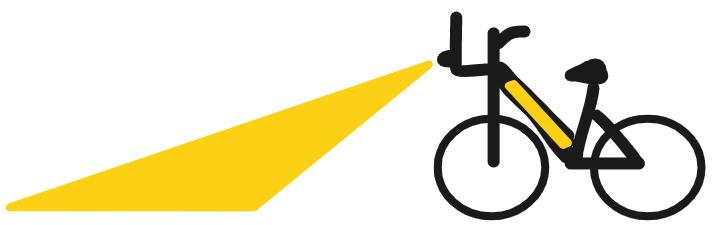}
\caption{Bicycle with lights.}
\label{fig:bike-with-lights}
\end{figure}

Lights line both sides of the bicycle frame, and a front light illuminates the path forward (Figure \ref{fig:bike-with-lights}). The lights stay steadily on when a bike is alone.  When a bike is joined by another bike that is equipped with the system, a swarm of two is formed, and the lights on both bikes begin to gently pulsate. The amplitude of the lights oscillates from high to low and back to high, in synchrony.  As other bikes come in proximity, their lights begin to pulsate synchronously as well, further growing the swarm and amplifying its visual effect.  

The system synchronizes swarms as they merge, as well as the momentary passing of bicycles. When any bike leaves the proximity of others, its lights return to their steady state. 

The effect is a unified pulsation of light, illuminating swarms of bikes as they move through the darkness.  This visual effect enhances their safety as well as their ability to attract more members to further grow the swarm.  Additionally, as a swarm grows and its perimeter expands, the reach of its radio messages expands as well, further enhancing its potential for growth.

While this paper focuses on the technical system that enables these swarms, we note that the swarming behavior that emerges can also be social.  Members of swarms may not know each other, but by riding in proximity, they collaboratively enhance the swarm's effects.

\section{Technical Description}

\subsection{A decentralized network of swarms}

In order to model swarms, we describe individual bikes as nodes in a decentralized network.  We consider swarms to be locally connected portions of the network, comprised of synchronized nodes.

The nodes synchronize by passing messages peer-to-peer and by running the same synchronization algorithm.  When nodes come within message-passing range of one another, they are able to connect and synchronize.  Two or more connected and synchronized nodes form a swarm.  When a node moves away from a swarm, and is no longer in range of message passing, it disconnects from that portion of the network, leaving the swarm. 

\begin{figure}[htbp]
\centerline{\includegraphics[width=1\columnwidth]{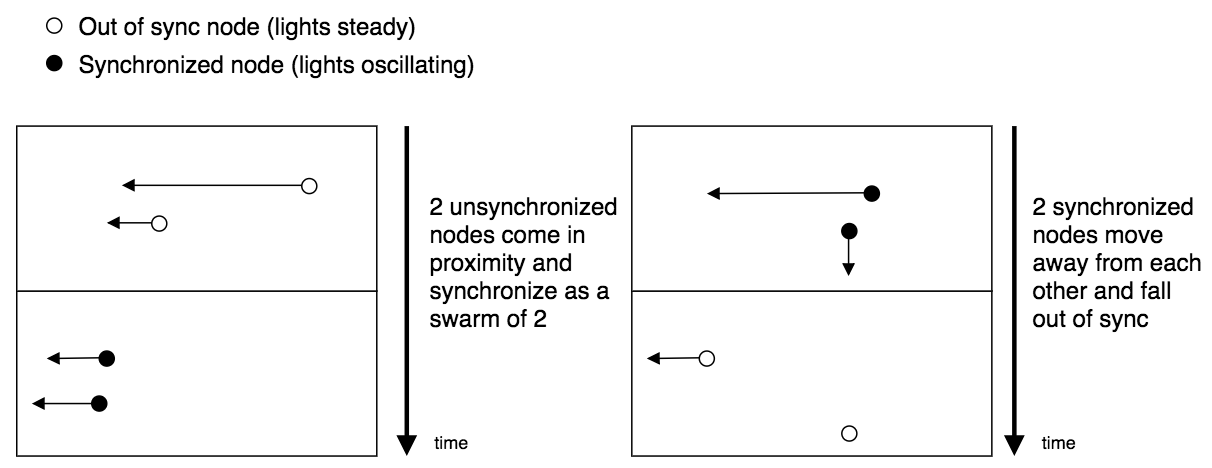}}
\caption{Nodes synchronize when near each other, and fall out of synchrony when they move apart.}
\label{fig:syn}
\end{figure}

The network's topology changes as nodes (bikes) move in or out of message passing range from one another, and connect or disconnect, and swarms thereby form, change shape, or dissolve (Figure \ref{fig:syn}).

\begin{figure}[htbp]
\centerline{\includegraphics[width=1\columnwidth]{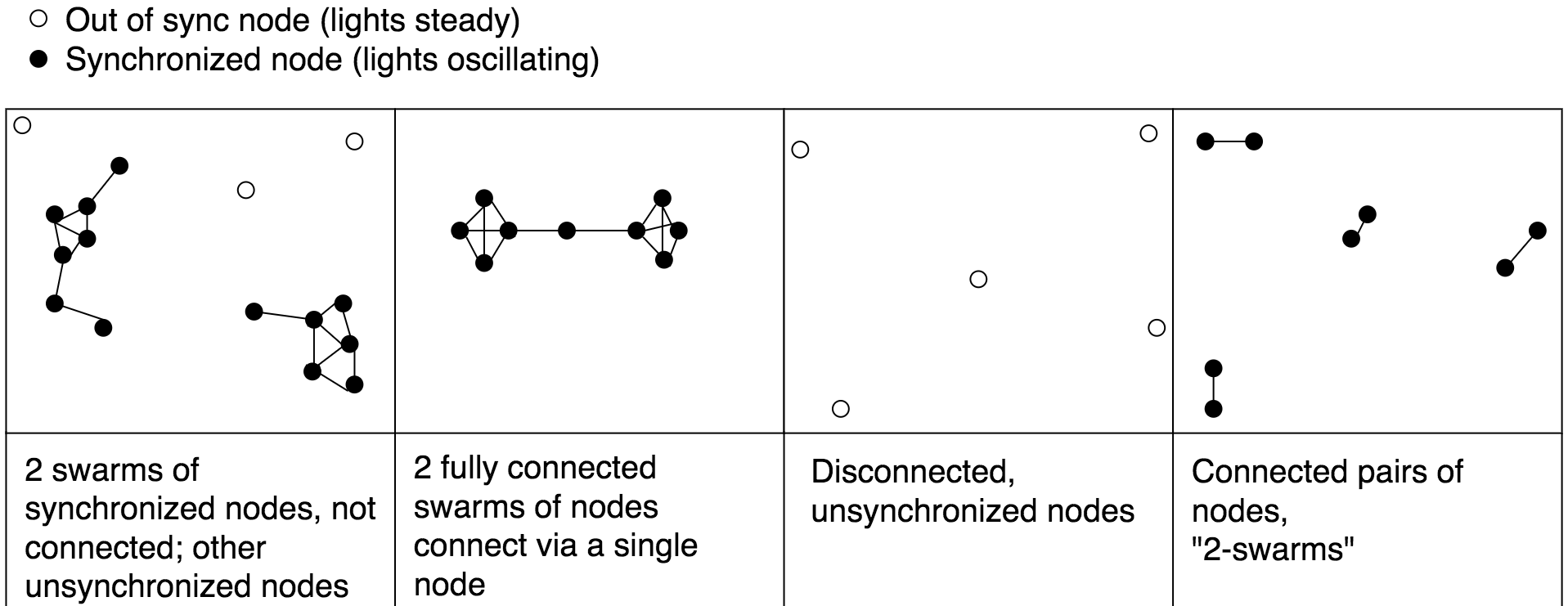}}
\caption{Examples of network states.}
\label{fig:netst}
\end{figure}

There may be multiple swarms of synchronized bikes in the city, with each swarm not necessarily in synchrony with another distant swarm. As such, the network of nodes may have a number of connected portions (swarms) at any given time, and these swarms may not be connected to one another (Figure \ref{fig:netst}).

\begin{figure}[htbp]
\centerline{\includegraphics[width=0.3\columnwidth]{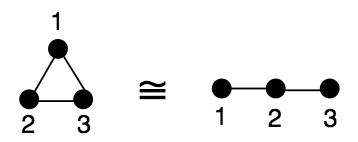}}
\caption{Synchrony of nodes in the network is transitive.}
\label{fig:trans}
\end{figure}

Our system exploits the transitive nature of synchrony: If \emph{node 1} is synchronized with \emph{node 2}, and \emph{node 2} is synchronized with \emph{node 3}, then \emph{node 1} and \emph{node 3} are synchronized as well. Since all nodes in a connected swarm are in synchrony with each other, a given node needs only to connect and synchronize with a single node in a swarm in order to synchronize with the entire swarm (Figure \ref{fig:trans}).

\begin{figure}[htbp]
\centerline{\includegraphics[width=0.9\columnwidth]{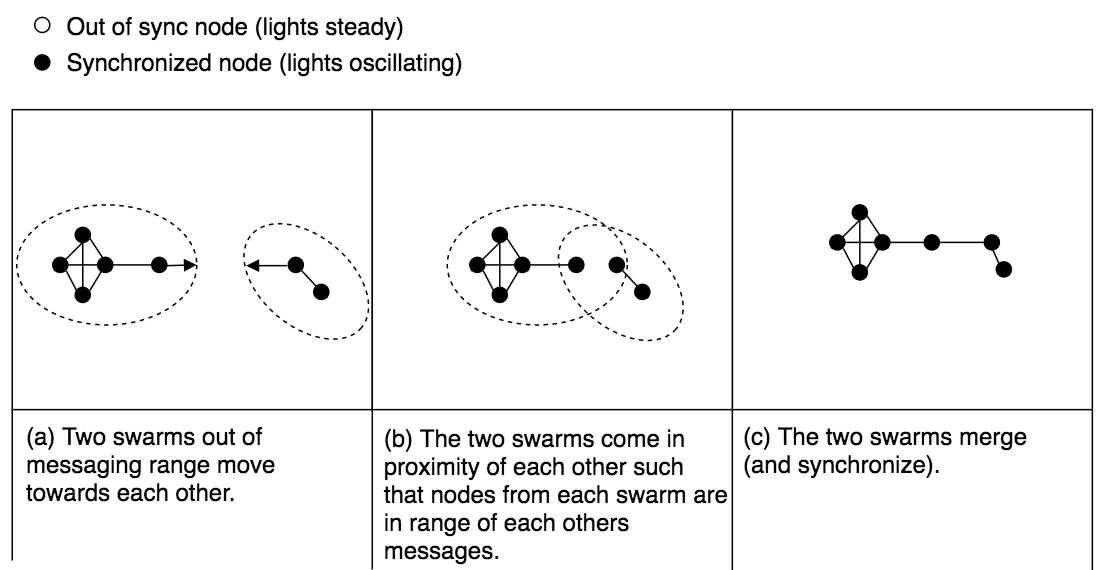}}
\caption{Two swarms come in proximity with each other and merge as one swarm.}
\label{fig:pros}
\end{figure}

When two synchronized swarms that are not in synchrony with each other come into proximity and connect for the first time, our message broadcasting protocol and synchronization algorithm facilitates their merge and transition towards a mutually synchronized state (Figure \ref{fig:pros}).

A feature of the message passing and synchronization protocol is that the nodes in the network need not be predetermined. New nodes can enter this decentralized network at any given time and immediately begin exchanging messages and synchronizing with pre-existing nodes.

\subsection{Nodes as oscillators}

The behavior of the nodes (bikes) that needs be synchronized is the timed pulsation of light. We can characterize this behavior by describing a node as an oscillator, similar to simple oscillators modeled in elementary physics. Nodes have two states:
\begin{enumerate}
\item Synchronized: the node's behavior is periodic and synchronized with another node.
\item Out of sync: the node's behavior remains steady; the node is not in communication with other nodes.
\end{enumerate}

All nodes share a fixed period, $T$.  When a node is in a state of synchrony, its behavior transitions over time, $t$, until $t=T$, at which point it returns to its behavior at time $t=0$. 

\begin{figure}[htbp]
\centerline{\includegraphics[width=0.5\columnwidth]{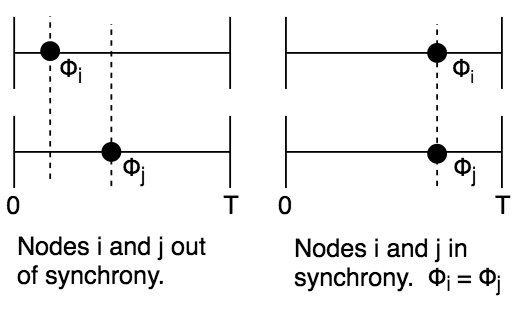}}
\caption{Out of sync nodes, and synchronized nodes.}
\label{fig:nodes-as-oscillators-in-and-out-of-sync}
\end{figure}

We denote the phase of node $i$ at time $t$ as $ \phi_i(t) $ such that $ \phi_i(t) \in [0, T]$ and the phases $0$ and $T$ are identical. When nodes are in synchrony, their phases are aligned. Thus for two nodes, node $i$ and node $j$, to be synchronized, $\phi_i(t) = \phi_j(t)$ (Figure \ref{fig:nodes-as-oscillators-in-and-out-of-sync}).

When a node is out of sync (i.e. the bike is not in proximity of another bike and therefore not exchanging messages with other bikes), then it ceases to act as an oscillator.  When out of the synchronous state, the node's phase remains stable at $\phi=0$.

\subsection{Phase and light}

\begin{figure}[htbp]
\centerline{\includegraphics[width=0.4\columnwidth]{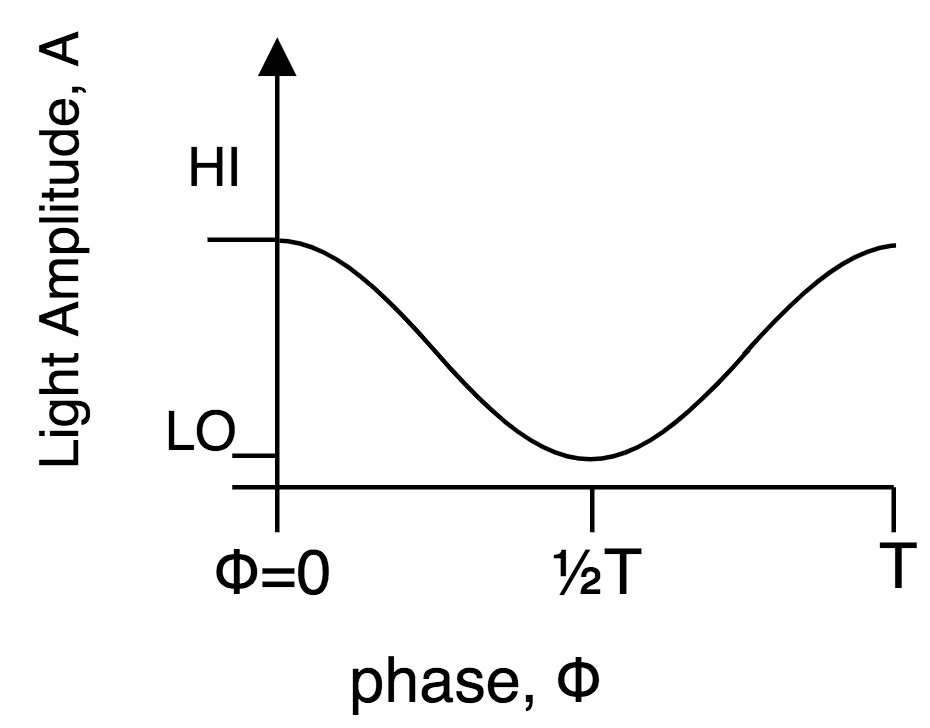}}
\caption{Graph of $f_A(\phi)$}
\label{fig:phi}
\end{figure}

The pulsating effect of a bike node's light is the decay and growth of the light's amplitude over the node's period, $T$.  The amplitude of the light is a function of the node's phase: $A = f_A(\phi)$ (see Figure \ref{fig:phi}).  
We denote the highest amplitude for the light as $HI$, and the lowest as $LO$\footnote{In our implementation, the amplitude of light does not reach as low as $0$ $(LO > 0)$. This decision was made due to our desired aesthetics and user experience.}, such that:

\begin{equation}
\label{eq:lightphase}
\begin{split}
f_A(0) = f_A(T)= HI \\ 
f_A(\tfrac{T}{2})=LO
\end{split}
\end{equation}

\begin{figure}[htbp]
\centerline{\includegraphics[width=0.6\columnwidth]{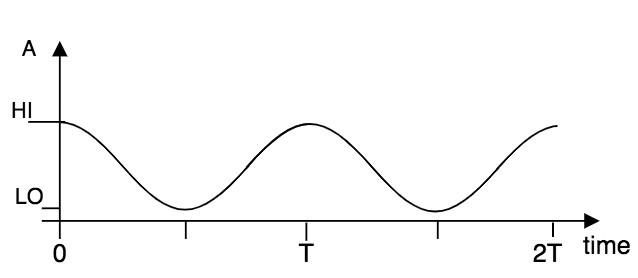}}
\caption{Amplitude, $A$, plotted as a function of relative time, $t$, for a node in the synchronized state.}
\label{fig:phase-and-light-in-sync-time-graph}
\end{figure}

When a node is in the synchronized state, and its phase oscillates, $\phi(t)\in[0, T]$, the amplitude of its light can be plotted as a function of time, $t$ (Figure \ref{fig:phase-and-light-in-sync-time-graph}).  
Note that nodes do not share a globally synchronized clock, so time $t$ is relative to the node. Without loss of generalization, we plot $t=0$ as when the given node enters a state of synchrony.

\begin{figure}[htbp]
\centerline{\includegraphics[width=0.6\columnwidth]{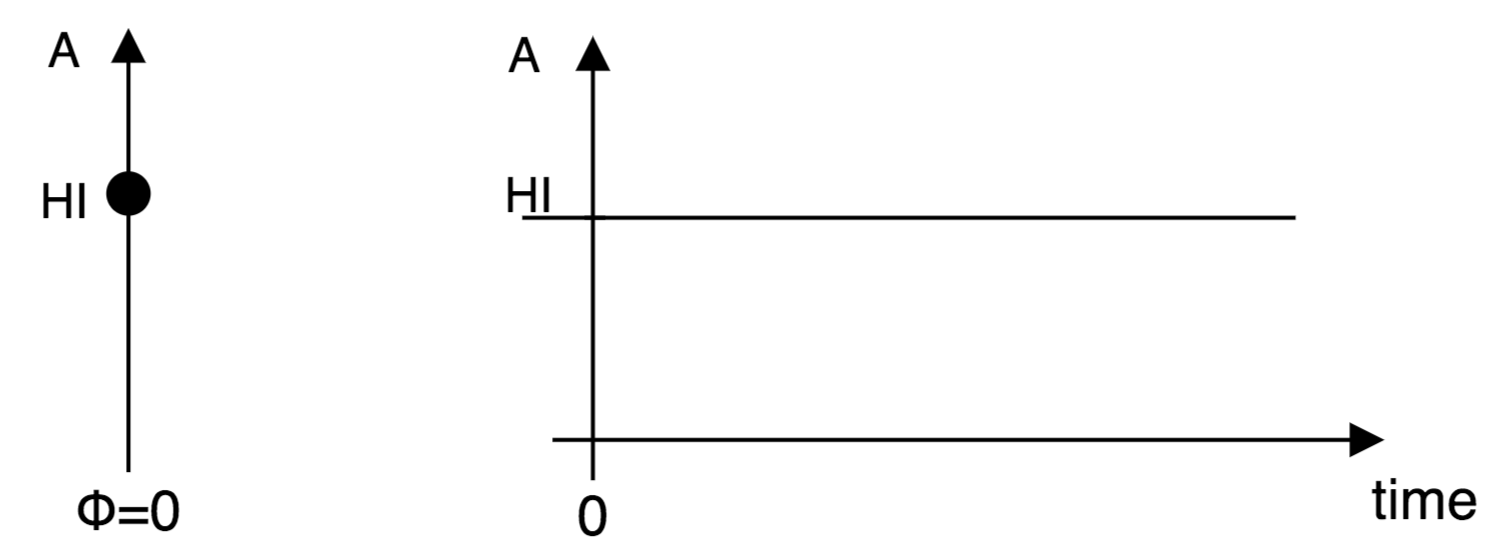}}
\caption{Amplitude, $A$, plotted for a node in the out of sync state.}
\label{fig:phase-and-light-out-of-sync}
\end{figure}

When a node is in the out of sync state, the value of its phase  $\phi$, is steady at $\phi=0$, so its light stays at the $HI$ amplitude.  $A = HI = f_A(0)$ (Figure \ref{fig:phase-and-light-out-of-sync}).

\begin{figure}[htbp]
\centerline{\includegraphics[width=0.6\columnwidth]{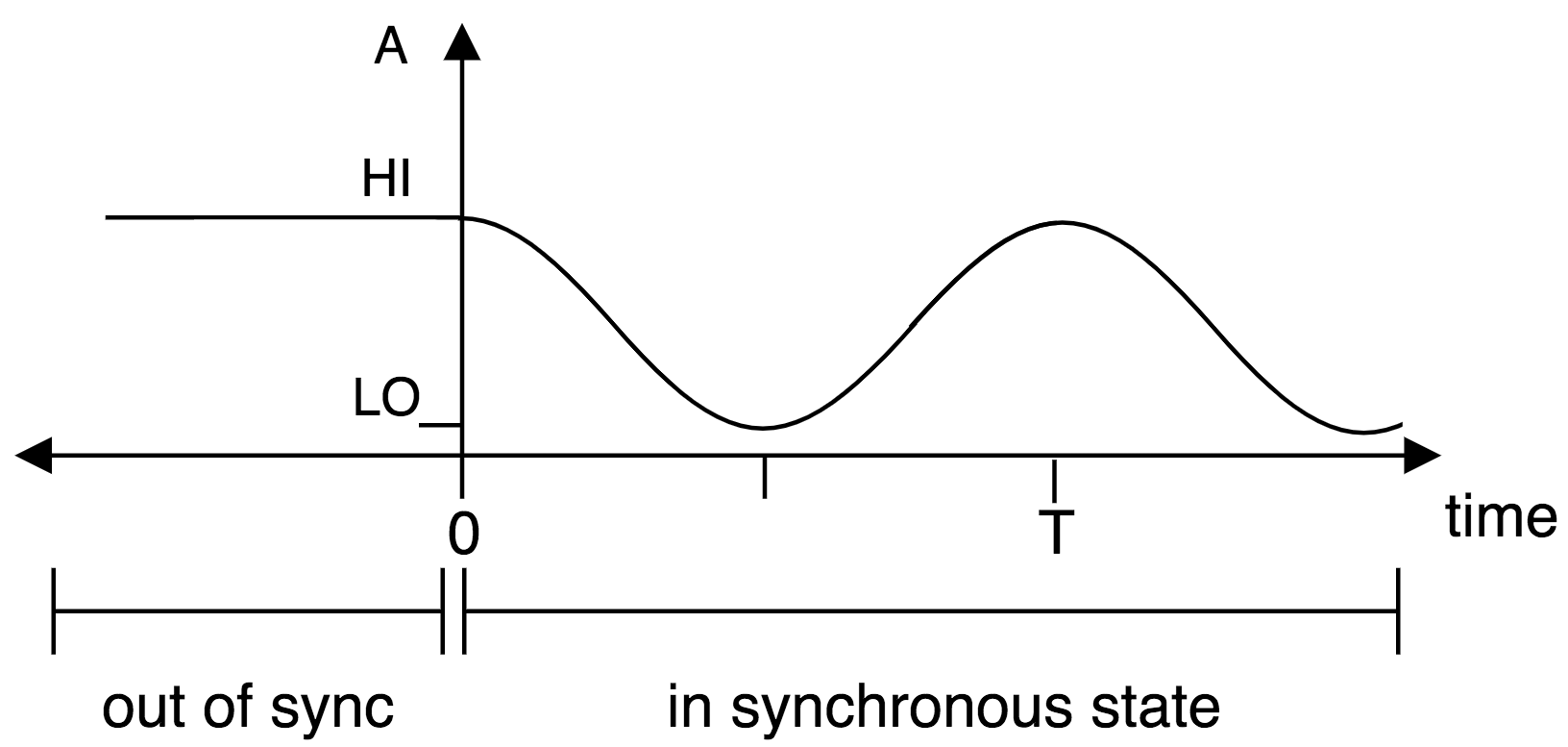}}
\caption{Amplitude, $A$, plotted as a function of time, for a node transitioning in to synchronous state.}
\label{fig:phase-and-light-out-and-into-sync}
\end{figure}

As soon as an out of sync node encounters another node and enters a state of synchrony, its phase begins to oscillate and the amplitude of its light transitions from $HI$ to $LO$ along the $f_A(\phi)$ path (Figure \ref{fig:phase-and-light-out-and-into-sync}).

\subsubsection{Implementation Notes}

For our bicycle lighting system we chose period $T=2200$ ms and chose $f_A(\theta)$ as a sinusoidal curve.

\begin{equation}
f_A(\phi) = \bigg[ \frac{cos(\phi * 2 \pi)}{T} + 1 \bigg] * \bigg[\dfrac{(HI - LO)}{2} \bigg] + LO
\end{equation}

We visually tested a variety of period lengths and functions.  We chose the combination that best produced a gentle rhythmic effect that would be aesthetically pleasing and noticeable, yet not distracting to drivers.

\begin{figure}[htbp]
\centerline{\includegraphics[width=0.33\columnwidth]{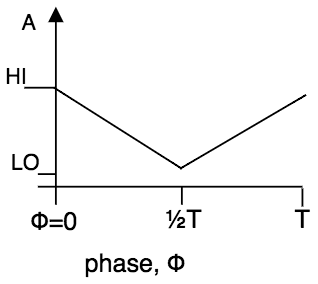}}
\caption{Graph of $f_A(\phi)$ as a piecewise linear function.}
\label{fig:piecewise_linear_f}
\end{figure}

We also considered $f_A(\phi)$ as a piecewise linear function (Figure \ref{fig:piecewise_linear_f}). For a slightly different effect, one might choose any other continuous function such that equations \ref{eq:lightphase} hold. As long as the period, $T$, is the same as other implementations, the nodes can synchronize.

\begin{equation}
f_A(\phi) =
\begin{cases}
HI - k * \phi, & \text{when  } \phi < \frac{T}{2} \\
LO + k *(\phi - \frac{T}{2}), & \text{when   } \phi \geq \frac{T}{2} 
\end{cases}
\end{equation}

\section{Message Broadcasting and Synchronization}

\subsection{Protocol}

Nodes maintain anonymity by communicating information pertaining only to timing over a broadcast and receive protocol. Synchronization is coordinated by a simple set of rules that govern how nodes handle messages received.

\subsubsection{Broadcasting messages}

The messages broadcast by a node are simply integers representing the node's phase, $\phi$, at the time of broadcasting, $t$,  i.e. nodes broadcast $\phi(t)$. Nodes in the out of sync state broadcast the message of $0$ (zero), as $\phi=0$ for out of sync nodes.

\subsubsection{Receiving messages}
Nodes update their phase values to match the highest phase value of nearby nodes. When a message is received by a node out of the synchronous state, the phase represented in the message, $\phi_m$, is necessarily greater than or equal, $\phi_m \geq \phi$, to the out of sync node's phase value of $\phi=0$.  The out of sync node then sets its phase to match the phase in the received message, $\phi = \phi_m$, and enters a state of synchrony. Its phase then begins to oscillate from the value of $\phi_m$, and bike lights pulsate in synchrony.

\begin{figure}[htbp]
\centerline{\includegraphics[width=0.5\columnwidth]{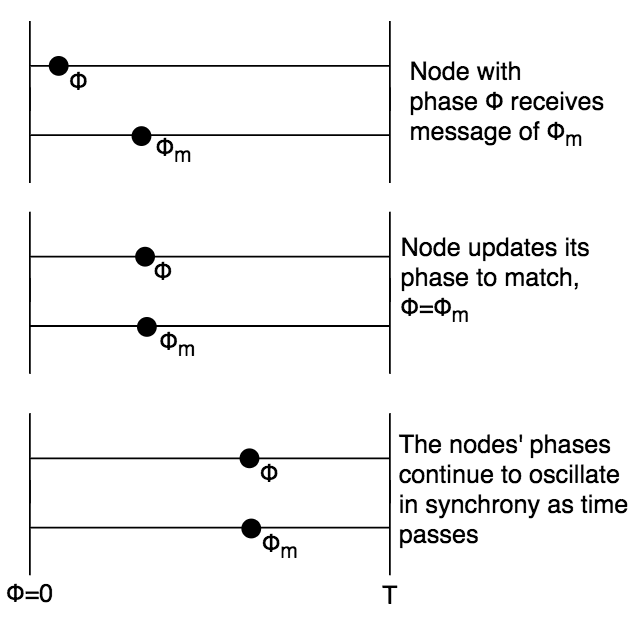}}
\caption{Node updates its phase value to match the phase value received in message.}
\label{fig:node_updates_phase_to_m}
\end{figure}

When a message is received by a node that is already in a state of synchrony, the node compares its own phase, $\phi$, to the phase represented in the received message, $\phi_m$.  If the node's phase value is less than the phase value in the received message, $\phi < \phi_m$, then the node updates its phase to match the received phase, $\phi=\phi_m$.  The node then continues in a state of synchrony, with its phase still oscillating, but now from the phase value of $\phi_m$.  The node is now in synchrony with the node that sent the message of $\phi_m$ (see Figure \ref{fig:node_updates_phase_to_m}).

\begin{figure}[htbp]
\centerline{
\includegraphics[width=0.9\columnwidth]{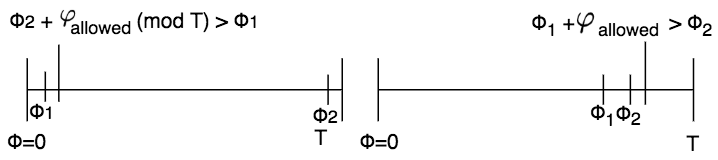}}
\caption{There is an allowed phase shift $\varphi_{allowed}$.}
\label{fig:allowed}
\end{figure}

There is an allowed phase shift, $\varphi_{allowed}$, to accommodate latency in message transit and receipt, and to keep nodes from changing phase more often than necessary (Figure \ref{fig:allowed}). Nodes do not update their phase to match a greater phase value if the difference between the phases is less than $\varphi_{allowed}$. For example, suppose \emph{node 1} has phase value $\phi_1$ and \emph{node 2} has phase value $\phi_2$, and $\phi_1 < \phi_2$. 
If $(\phi_1 + \varphi_{allowed}  > \phi_2)$ or $(\phi_2 + \varphi_{allowed} \mod{T} ) > \phi_1$ then \emph{node 1} does not update its phase to match $\phi_2$ upon receiving a message of $\phi_2$.  In our implementation, $\varphi_{allowed}$ is so small that the possible phase shift between the light pulsations of bike nodes is imperceptible.

Once a node updates its phase to match a greater phase received in a message, $\phi_m$, it then broadcasts its new phase. Nodes in range of this new message may have been out of range of the original message, but these nearby nodes can now all synchronize around the new common phase $\phi_m$. This simple protocol works as a mechanism for multiple swarms to merge and synchronize.

\begin{figure*}%
    \centering
    \includegraphics[width=0.65\textwidth]{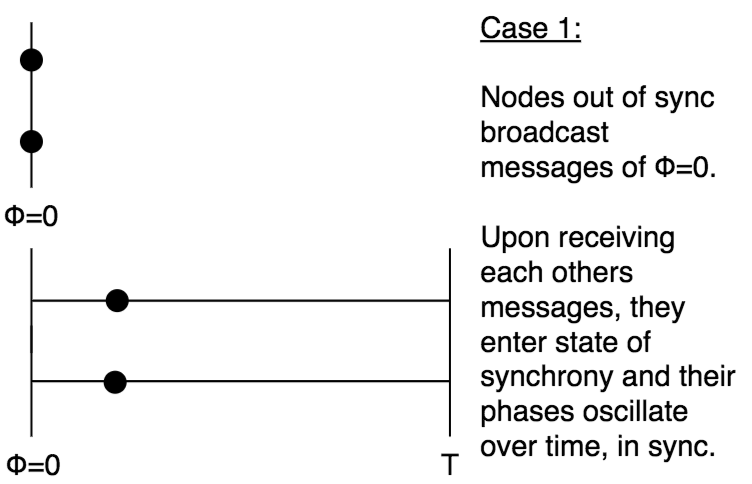} \\
    \vspace{1.2cm}
    \includegraphics[width=0.65\textwidth]{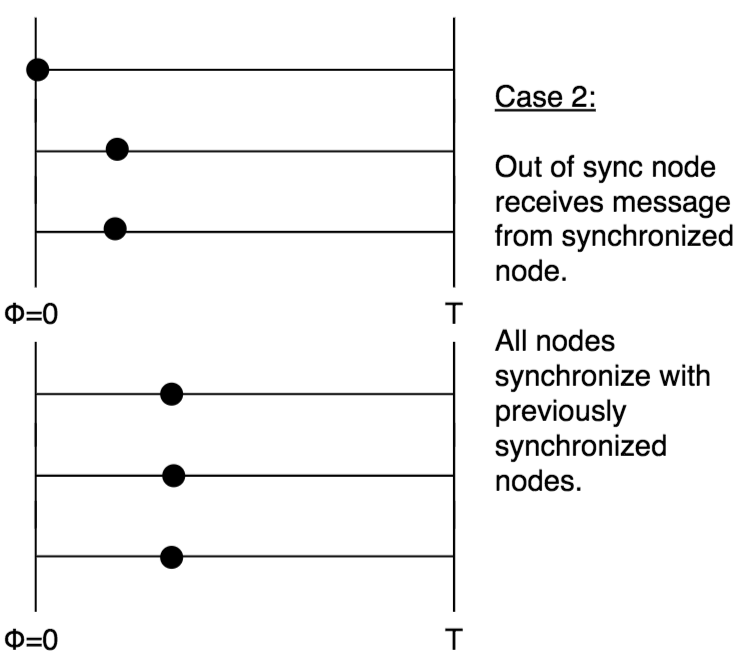} \\
    \vspace{1.2cm}
    \includegraphics[width=0.65\textwidth]{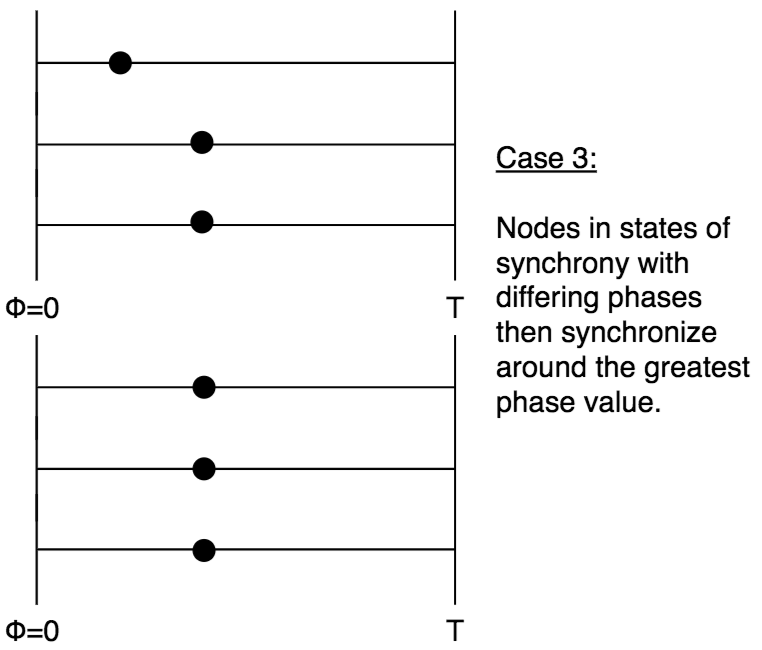}
    \caption{Scenarios of nodes receiving broadcast messages and updating their state of synchrony.}%
    \label{fig:sync_cases}%
\end{figure*}

Moreover, whenever nodes come in proximity of each other's messages, they will synchronize. Even when node $i$ with phase value $\phi$ receives message $\phi_m < \phi$ from node $m$, and node $i$ does not updates its phase to match $\phi_m$, node $i$ and node $m$ will still synchronize. Since they are in message passing range, node $m$ will receive the message broadcast by node $i$ of $\phi > \phi_m$, and node $m$ will then update is own phase to match $\phi$. Figure \ref{fig:sync_cases} illustrates various scenarios for receipt of the broadcast message.

Once nodes synchronize, minimal messages are required to keep them synchronized, as all nodes share the same period of oscillation, $T$. When enough time passes without a node receiving any messages, the node then leaves its synchronous state and returns to the out of sync state where its phase stays steady at $\phi=0$ (and its lights cease to pulsate).

Consider the cases of Figure \ref{fig:sync_cases} where nodes come in range of each other's messages and synchronize.
We let the reader extend these small examples to the larger network topology of nodes previously provided.

The broadcast messages are minimal, and the synchronization rule set simple, and we consider this simplicity a feature. We demonstrate its implementation as an algorithm.

\subsection{Algorithm}
The implementation of our algorithm used for our working prototype is provided open source 
\footnote{\url{https://github.com/aberke/city-science-bike-swarm/tree/master/Arduino/PulseInSync}}.

Nodes execute their logical operations through a continuous loop. Throughout the loop, they listen for messages, as well as update their phase as time passes. 
Algorithm \ref{algo:update_phase} and Algorithm \ref{algo:loop} outline the loop operations.

This simple protocol and algorithm offer the following benefits across the network, with the only requirements being that all nodes in the network run loops with this same logic, and share the same fixed period.

\begin{itemize}
\item Nodes need not share a globally synchronized clock in order to synchronize their phases. Time can be kept relative to a node.
\item Nodes need not share any metadata about their identity, nor need to know any information about other nodes, in order to synchronize. Unknown nodes can arbitrarily join or leave the network at any time while the network maintains its mechanisms for synchrony.
\end{itemize}

\begin{algorithm}
 \caption{Routine to update phase} \label{algo:update_phase}
 \begin{algorithmic}[1]
 \STATE currentTime $\gets$ getCurrentTime() 
  \IF {node is inSync} 
    \STATE timeDelta $\gets$ currentTime - lastTimeCheck
    \STATE phase $\gets$ (phase + timeDelta) \% period
  \ELSE
    \STATE phase $\gets$ 0
  \ENDIF 
  \STATE lastTimeCheck $\gets$ currentTime
  \RETURN{}  phase
 \end{algorithmic}
 \end{algorithm}

 \begin{algorithm}
 \caption{Main loop} \label{algo:loop}
 \begin{algorithmic}[1]
 \STATE inSync $\gets$ FALSE 
 
 \IF{currentTime $-$ lastReceiveTime $<$ timeToOutOfSync}
      \STATE  inSync $\gets$ TRUE
   \ENDIF
  \STATE phase $\gets$ updatePhase()
  \STATE phaseM $\gets$ receive()
   \IF{phaseM not null}
	\STATE lastReceiveTime $\gets$ getCurrentTime()
	\STATE phase $\gets$ updatePhase()
		\IF{ (phase $<$ phaseM) \textup\& (computePhaseShift(phase, phaseM) $<$ allowedPhaseShift)}
		\STATE phase $\gets$ phaseM + expectedLatency
		\STATE lastTimeCheck $\gets$  lastReceiveTime
		\ENDIF	
	 \ENDIF
\STATE broadcast(phase)
\STATE phase $\gets$ updatePhase()
\STATE updateLights(phase)
\end{algorithmic}
\end{algorithm}
 
 \subsection{Addressing scheme}

A requirement of the system is that any two nodes must be able to communicate upon coming in proximity of one another, without knowing information about the other beforehand.
Moreover, any new node that enters an existing network must be able to immediately begin broadcasting and receiving messages to synchronize with pre-existing nodes in the network.  Thus the challenge is to accomplish this communication without nodes sharing identities or addresses.  Because these nodes are broadcasting and receiving messages over radio, nodes cannot simply all broadcast and receive messages on the same channel, or else their messages will conflict and communication will be lost.

Methods have been developed to facilitate resource sharing among nodes in a wireless network such as our network of bike nodes (e.g. TDMA implementations \cite{miao2016fundamentals}). These methods are designed to avoid the problems of nodes sending messages on the same channel at conflicting times by coordinating the timing at which messages are sent. The DESYNC algorithm \cite{degesys2007desync} even supports channel sharing across decentralized networks of nodes that do not share a globally synchronized clock (such as our network), by nodes monitoring when other messages are sent, and then self-adjusting the time at which they send messages, until gradually the nodes send their messages at equally spaced intervals.

These strategies are not as well suited for our network of bike nodes, because its topology continuously changes (as new bikes join or leave the network, and as bikes pass each other, or collect at stoplights, or go separate ways), and nodes need to exchange messages as soon as they enter proximity of each other. In addition, immediately after a node updates its own phase to match a phase received in a message, it must broadcast its phase so that other nearby nodes can resynchronize with it. This immediate resynchronization would be hindered by a resource sharing algorithm that required a node to wait its turn in order to broadcast a message.

Bike nodes should be able to continuously listen for messages sent by other nodes, and be able to broadcast messages at any time. We designed and use an addressing scheme to handle these requirements. The scheme exploits the fact that when multiple nodes are in proximity of each other, the messages they broadcast are often redundant: When nodes are in message passing range, they synchronize and the messages they then broadcast contain the same information about their shared phase. 

In our addressing scheme, we allocate $N$ predetermined addresses, which we number as \emph{address 1}, \emph{address 2}, \emph{address} $3,\dots,$ \emph{address} $N$. All nodes in the network know these common addresses in the same way they all know the common period, $T$.

We also consider our nodes as numbered: 
\begin{itemize}
    \item[] \emph{node 1}, \emph{node 2}, \emph{node 3}, $\dots$
\end{itemize}

Each node uses one of the $N$ \emph{addresses} to broadcast messages, and listens for messages on the remaining $N-1$ \emph{addresses}:

\begin{itemize}
  \item \emph{node} $i$ broadcasts on \emph{address} $i$, 
  \item \emph{node} $i$ listens on \emph{address} $i+1 \mod N$, \emph{address} $i+2 \mod N$,\dots, \emph{address} $i + (N-1) \mod N$
\end{itemize}

For example, \emph{node 1} broadcasts on \emph{address} $1$, while \emph{node 2} broadcasts on \emph{address} $2$. Since \emph{node 1} also listens on \emph{address 2}, and \emph{node 2} listens on \emph{address 1}, the two nodes can exchange messages without conflict.

Nodes determine their own node numbers by randomly drawing from a discrete uniform distribution over $\{1, 2, 3, $\dots$, N\}$, such that node $i$ had a $\tfrac{1}{N}$ chance of choosing any $i \in \{1, 2, 3, $\dots$, N\}$. 

When a node in the out of sync state comes in proximity of another node, there is a small ($\tfrac{1}{N}$) probability that the nodes share a node number and therefore will not be able to exchange messages. To overcome this issue, nodes in the out of sync state regularly change their node numbers by redrawing from the discrete uniform distribution and then re-configuring which addresses they broadcast and listen on based on their node number. This change allows two nearby nodes with conflicting node numbers and addresses to get out of conflict.

If a node encounters multiple synchronized nodes, it needs only to have a non-conflicting node number with one of them in order to synchronize with all of them, since the synchronized nodes share and communicate the same phase messages.

\subsubsection{Discussion of alternatives}

We also considered an alternative synchronization scheme that would allow all nodes to share one common address to broadcast and receive messages. In this simpler alternative, nodes only broadcast messages when their phase is at $0$. (Nodes in the out of sync state broadcast at random intervals). Upon receiving such a message, a node sets its own phase to $0$ and enters a state of synchrony with the message sender. Simplifying the message protocol in this way circumvents the issue of nodes sending messages over a shared channel and their messages conflicting. If two nodes do happen to send a message at the same time, then they must already be synchronized (they share a phase of $0$ at the time of sending). Any other node in proximity that receives this broadcast will synchronize to phase $0$ and then also broadcast its messages at the same time as the other nodes.

This message passing protocol has been studied and modeled in relation to pulse-coupled biological oscillators where the oscillation is episodic rather than smooth \cite{mirollo1990synchronization}. Examples include the flashing of a firefly, or the chirp of a cricket, where instead of the system interacting throughout the period of oscillation, there is a single ``fire'' (e.g. flash or chirp) event that occurs at the end of the period.

\begin{figure}[htbp]
\centerline{\includegraphics[width=1.0\columnwidth]{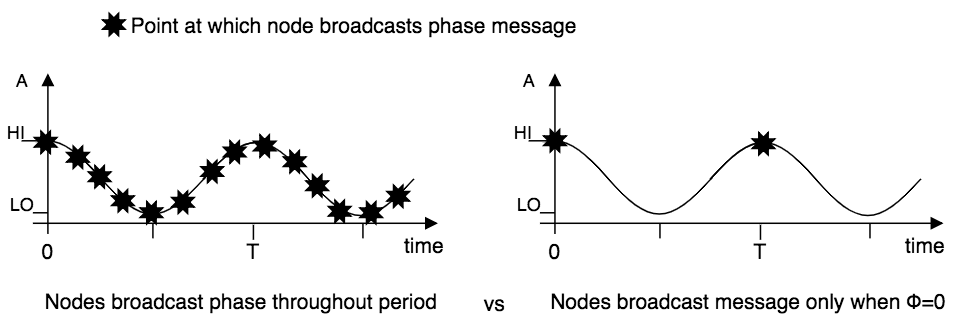}}
\caption{The timing of message broadcasts in our synchronization scheme versus episodic message broadcasts in the simplified synchronization scheme.}
\label{fig:messaging-protocol-broadcast-on-fire}
\end{figure}

This simplified synchronization scheme works well for discrete episodic events among oscillators, and while it could work for our bike nodes, we chose not to use it because our bike nodes have a continuous behavior (Figure \ref{fig:messaging-protocol-broadcast-on-fire}). Because they update the amplitude of their light continually throughout their phase, synchronizing at phase $0$ is as important as synchronizing at any other phase value. Moreover, this simplified messaging protocol would make the time to synchronization longer, dependent on the length of the period, $T$. Two nodes that come in to proximity for the first time but that are already oscillating with phases that are out of synchrony with each other would not have the opportunity to synchronize until one of their phases reaches $0$ again.

\subsection{Faulty and malicious nodes}

We note the unlikely case of faulty nodes, which broadcast messages to the bike swarm network without following the same protocol as other nodes.  These faults may occur because one bike's system breaks or was badly implemented, or may be due to malicious actors. These faulty nodes can destabilize the synchronization of nearby swarms.  However, the issue will be spatially isolated to the swarms within broadcasting range of the faulty node, while the rest of the network continues to function successfully.

 \section{Circuitry, Prototype Fabrication, and Tests}

Our system design includes an integrated circuit.  The circuit connects a low-cost radio to broadcast and receive the protocol messages, a microcontroller programmed to run our algorithm, and lights controlled by the microcontroller.  The radio transmission range is limited by design in order to control swarm membership to only include nearby nodes (bikes).

Our prototypes use nRF24 \cite{nrf24series-nordic-semiconductor} radio transceivers to broadcast and receive the protocol messages without necessitating individual nodes to pair. The nRF24 specification allows for software control of transmission range, which is used to constrain the spatial distance between connected nodes. The Arduino Nano \cite{wiki:Arduino} microcontroller was selected to run the synchronization protocol and algorithm.  The other components in our circuit were used for the management of power and the pulsation of lights.  The circuit schematic and Arduino code are open source.
\footnote{\url{https://github.com/aberke/city-science-bike-swarm}}.

We implemented and tested our system for urban mobility swarms by fabricating a set of 6 prototypes.  The prototypes strap on and off bicycles from a city bike-share program, and we rode throughout our city with them over a series of three nights.  We tested various scenarios of bikes forming, joining, and leaving swarms, as well as swarms passing, and swarms merging, as shown in video footage that is available online: 
\url{https://youtu.be/wUl-CHJ6DK0}.
Also available online is detailed photographic documentation of the prototype development and deployment:
\url{https://www.media.mit.edu/projects/bike-swarm}.

\section{Conclusion}

We designed a system for the urban environment that draws from swarming behavior exhibited in the natural environment.  In this paper we presented urban mobility swarms as a means to promote the use and safety of lightweight, sustainable transit.  We described and demonstrated a system for their implementation, with a radio protocol, synchronization algorithm, and tested prototypes.

The prototypes we designed are specific to synchronizing the lights of nearby bicycles in the dark.  Riders within swarms collaboratively amplify the swarm's effect and collective safety, yet coordination and formation of swarms requires no effort from the riders.  The riders are automatically inducted into ad-hoc swarms when in proximity due to our simple, yet powerful system design.

The system we implemented can be easily extended and applied to transit options beyond bicycles.  More generally, our system treats individual riders as nodes in a decentralized network, and coordinates swarms as connected portions of the network, with a peer-to-peer message protocol and algorithm.  Our design accommodates a dynamically changing network topology, as necessitated by the nature of an urban mobility network in which individuals are constantly moving, joining the network, or leaving altogether.  Furthermore, the features of our decentralized design afford its flexible and secure implementation.  There is no global clock and nodes communicate with minimal radio messages without sharing metadata, allowing new nodes to immediately coordinate with the system while maintaining an individual's privacy.

Moreover, our system can be deployed at scale, which we demonstrated by implementing it with simple, low-cost circuit and hardware components, and by testing with bikes from a city bike-share. Bike-share programs manage fleets of bikes distributed across cities and could deploy the system at city-scale.  The system can be integrated into bicycles, or strapped on and off, as riders typically do with bike lights.  Once deployed, our modular hardware and decentralized system design allows arbitrary bikes to form or further grow a network with ease.
 
\bibliographystyle{splncs}
\bibliography{references}

\end{document}